# Observation of Ultraslow Group Velocity of Light in a Solid

**A.V. Turukhin[1], V.S. Sudarshanam, and M. S. Shahriar**

Research Laboratory of Electronics, Massachusetts Institute of Technology,

77 Massachusetts Avenue, Cambridge, MA 02139

**J. A. Musser**

Texas A&M University, College Station, TX 77843

**P.R. Hemmer**

Air Force Research Laboratory, Sensors Directorate, Hanscom AFB, MA 01731

**Abstract:** We report ultraslow group velocities of light in a solid. Light speeds as slow as 45 m/s were observed, corresponding to a group delay of 66 μs in a 3-mm thick crystal. Reduction of the group velocity is accomplished by using a sharp spectral feature in absorption and dispersion, produced by a Raman excited spin coherence in an optically dense Pr doped $Y_2SiO_5$ crystal.

[1] Corresponding author: Phone: 617-253-3072 Fax: 801-469-1832 Email: alexey@mit.edu

Since the first observation of ultraslow light[1], there has been substantial interest in its potential applications. For example, it was proposed that slowing the group velocity of a laser pulse down to the speed of sound in a material can produce strong coupling between acoustic waves and the electromagnetic field[2]. The resultant giant nonlinearity obtained by this method might be utilized for efficient multi-wave mixing and quantum nondemolition measurements[3], as well as for novel acousto-optical devices. A carefully controlled, slow group velocity of light might even allow a very efficient nonlinear interaction between laser pulses of extremely low (down to a single photon) energy[4]. These effects can be used to create quantum entanglement between single photons without an ultrahigh finesse cavity and therefore are of great interest for quantum information processing. Slow light is also of interest because it serves as a useful metric to compare the relative merits of dissimilar nonlinear optical interactions. This is especially important for applications that make use of large optical dispersive properties of EIT[5-9] for the detection of small phase shifts, such as in very sensitive magnetometers.[10]

Initially, ultraslow light was observed in an ultra cold atomic vapor using the sharp dispersion caused by electromagnetically induced transparency (EIT)[11]. EIT, which is related to coherent population trapping[12], uses an intense electromagnetic field is used to modify the absorption coefficient and refractive index seen by a weak probe so that an optically dense media can be made nearly transparent to light at its resonance frequency. In addition, the sharp dispersion of EIT leads to a slow group velocity for the propagation of light, tuned to the frequency of the transparency peak. At first it was believed that only a material with a small inhomogeneous broadening could be used to achieve ultraslow light, but subsequent experiments employing a hot vapors[5,8,13] were soon shown to work better than expected.[5,8] Still, for a number of applications, the need to compensate for Doppler shifts in a hot vapor by using nearly co-propagating laser beams is inconvenient. Finally, for potential technological device applications of ultraslow light, there is no substitute for solids.

Recently, we reported near 100% efficient EIT in a $Pr^{3+}$ doped $Y_2SiO_5$ (Pr:YSO) crystal[14] wherein transparency of probe field was demonstrated at line center in an optically thick sample. At low temperatures, this "dark resonance" has a width on the order of 10's kHz. This potentially makes it suitable for the direct measurement of ultra long optical group delays because the group delay time for a pulse that propagates one attenuation length is approximately given by to the inverse width of the EIT peak[8] assuming 100% efficient EIT. We would like to note that Pr:YSO and many rare-earth doped crystals have properties similar to both hot and cold atomic vapors. In particular, both Pr:YSO and ultra cold vapors, have the advantage that there is no motional diffusion of atoms, whereas Pr:YSO and hot vapors have the common advantage that the light speed can be slowed to well below the sound speed in the medium.

Our crystal was supplied by Scientific Materials, Inc. and consisted of 0.05 at% Pr doped YSO in which $Pr^{3+}$ substitutes $Y^{3+}$. The crystal had a thickness of 3 mm in the light propagation direction. To generate the EIT peak, we used the $^3H_4 \rightarrow {}^1D_2$ optical transition with a central frequency of 605.7 nm. The relevant energy level diagram is presented in Figure 1. Here, the coupling and the probe fields, with frequencies $\omega_C$ and $\omega_P$ respectively, create a coherence the between ground states $^3H_4$ ($\pm 3/2 \leftrightarrow \pm 5/2$), while the repump field with frequency $\omega_R$ partially refills the spectral holes burned by the coupling and the probe fields. This partial refilling or anti-hole also provides a narrow effective optical inhomoheneous linewidth for the Raman transitions. The experimental arrangement is similar to that used in Ref. 14. This time, we used a COHERENT 899-21 single mode ring dye laser pumped by an INNOVA 300C argon laser. The dye laser was continuos wave with a laser jitter measured to be about 0.5 MHz. All the laser fields shown in Figure 1 were derived from the dye laser output using acousto-optic frequency shifters. This greatly relaxes dye laser frequency stability requirements since the resonant Raman interaction is insensitive to correlated laser jitter. To match Figure 1, the coupling, repump, and probe beams were downshifted from the original laser frequency by 288.3, 311.0, and 298.5 MHz respectively. To generate the probe beam absorption spectra, the probe frequency was scanned around the 10.2 MHz Raman transition frequency while the frequencies of the coupling and the repump beams were held fixed. A set of compensating galvos was used to correct for the optical alignment changes that arise when the acousto-optic modulators are tuned. The intersection angle of the coupling and probe beams at the crystal was about $4^o$ in the plane of the optical table. The intersection angle of the repump beam was also $4^o$ but out of the plane of the optical table. This geometrical arrangement allowed us to reduce the effects of scattered light from the coupling and the repump beams. To further increase the signal to noise ratio, the probe beam was modulated and the transmitted signal was detected by standard phase-sensitive detection techniques. All laser beams were linearly polarized and focused into the crystal by a 150 mm focal length lens, producing a spot with a diameter of about 100 $\mu$m. The beams were polarized in a common direction that could be rotated by a half-wave plate to maximize absorption for a given orientation of the crystal. During the experiment, the sample was maintained at a temperature of 5.5 K inside a helium flow JANIS cryostat.

The dye laser was locked near the center of the inhomogeneously broadened absorption line at 605.7 nm. The intensities of the probe and repump fields were held fixed during the experiment at 0.1 and 1.6 W/cm$^2$, respectively. The background absorption was determined by the anti-hole, which was created by the repump beam, and superimposed on the much broader, saturated hole burned by the pump beam. The width of anti-hole was found to be about 0.5 MHz and was defined by the dye laser frequency jitter. The width of the saturated hole was not determined due to the limited 50 MHz bandwidth of the acousto-optic modulators, but appeared to be much larger than 50 MHz. Due to the limited intensity of the coupling field, the maximal transparency was found to be about 50%. Figure 2a shows a representative absorption spectrum of the probe field obtained at a coupling field intensity of 105 W/cm$^2$. The spectrum was recorded using phase in-sensitive lock-in detection. The measured FWHM of EIT peak was around 60 kHz for the chosen range of coupling field intensities. This is much smaller than the laser jitter of 0.5 MHz and is in agreement with the inhomogeneous linewidth of the spin transition as measured by optically detected nuclear magnetic resonance (ODNMR). The peak absorption of the probe near the center of the anti-hole line was about 90% and could be adjusted by changing the intensities of the coupling and the repump fields. Blocking either the coupling beam or the repump beam resulted in nearly 100% transparency of the probe beam because of spectral hole burning.

The temporal retardation of the amplitude modulated probe beam tuned to the EIT peak manifests itself as a strong shift in the modulation phase that can be measured with a lock-in amplifier. Figure 2b and 2c show the absorption spectra of the probe beam for two different lock-in amplifier phases. Both spectra were recorded at a coupling beam intensity of 105 W/cm$^2$ and a probe beam amplitude modulation frequency of 6 kHz. Figure 2b shows the absorption spectrum recorded with the lock-in amplifier phase adjusted to suppress the broad absorption, and Figure 2c demonstrates the same absorption spectrum with the lock-in amplifier phase adjusted to suppress the EIT peak. Comparison of these two spectra clearly shows a strong dependence of the phase of modulation of the transmitted signal on the detuning from the EIT peak. To measure the group delay, the probe field was modulated with a square wave at a frequency that was varied in the range of 3 – 6 kHz. These low modulation frequencies assure that pulse spectrum is contained within the EIT peak. The group time delay was calculated based on the modulation frequency and the phase shift between the transmitted probe signal with and without the EIT-producing coupling field. Figure 3 presents the measured phase shift as a function of modulation frequency for the coupling beam intensity of 105 W/cm$^2$. The group delay time for this coupling beam intensity was found to be 39.6 μs from the linear fit of the data. This result is in good agreement with an estimate based on the equations in ref. 13 using the measured 62 kHz FWHM of the EIT resonance.

At low coupling beam intensities, the fraction of Pr ions that are pumped into the "dark state" increases with coupling laser power and therefore the depth of the transparency increases. This causes the group delay to increase. At higher intensities, the transparency begins to power broaden, reducing the sharpness of the dispersive feature. This reduces the observed group delay[5]. To show these saturation effects, the EIT peak width and amplitude were measured as a function of the coupling beam intensity and the results are shown in Figure 4. At intensities below 50 W/cm$^2$ the width of the EIT peak is defined by the inhomogeneous width of the ground state transition, and therefore it is a constant. Meanwhile the amplitude of the EIT peak increases linearly with the intensity. At high intensities above 65 W/cm$^2$ the width of the peak begins to increase with intensity due to power broadening, while the amplitude of the peak saturates. Comparing this data to Figure 5, which shows the group velocity and delay versus coupling beam intensity, it is apparent that there is a strong correlation between the EIT linewidths and amplitudes and the observed group velocities.

As can be seen in Figure 5, a group velocity of light as slow as 45 m/s is reached. Here, we assumed the length of an interaction zone to be 3 mm, which is the thickness of the crystal. However, due to the geometric configuration used, the interaction area where all three beams are overlapping is slightly shorter than the thickness of the crystal, so that the actual group velocities are slightly smaller than in Figure 5. In contrast to the slow light experiment in an optically dense hot vapor[13], the light velocity here is practically independent of the propagation distance since the intense coupling beam propagates with little attenuation. However, as shown in Figure 5 the dependence of the group velocity on the input coupling field intensity still permits fine tuning of the group delay so as to allow one to achieve phase-matched conditions in an experiment on stimulated Brillouin scattering[2] and/or to adjust group velocities of single photons for ultrastrong nonlinear interaction[4].

In conclusion, we demonstrated an ultraslow group velocity of light of 45 m/s in a rare-earth doped crystal, which is fundamentally limited only by the inhomogeneous broadening of the ground state transition. This first observation of ultraslow light in a solid is an key enabling step toward many potential applications of slow light to low-intensity nonlinear optics, quantum information processing, as well as to technological devices.

We acknowledge discussions with S. Ezekiel of Massachusetts Institute of Technology. The authors also wish to thank Mr. John Kierstead for the help with RF electronics. This study was supported by the Air Force Research Laboratory (contract #F19628-00-C-0074), ARO grant #s DAAG55-98-1-0375 and DAAD19-001-0177, and AFOSR grant # F49620-98-1-0313.

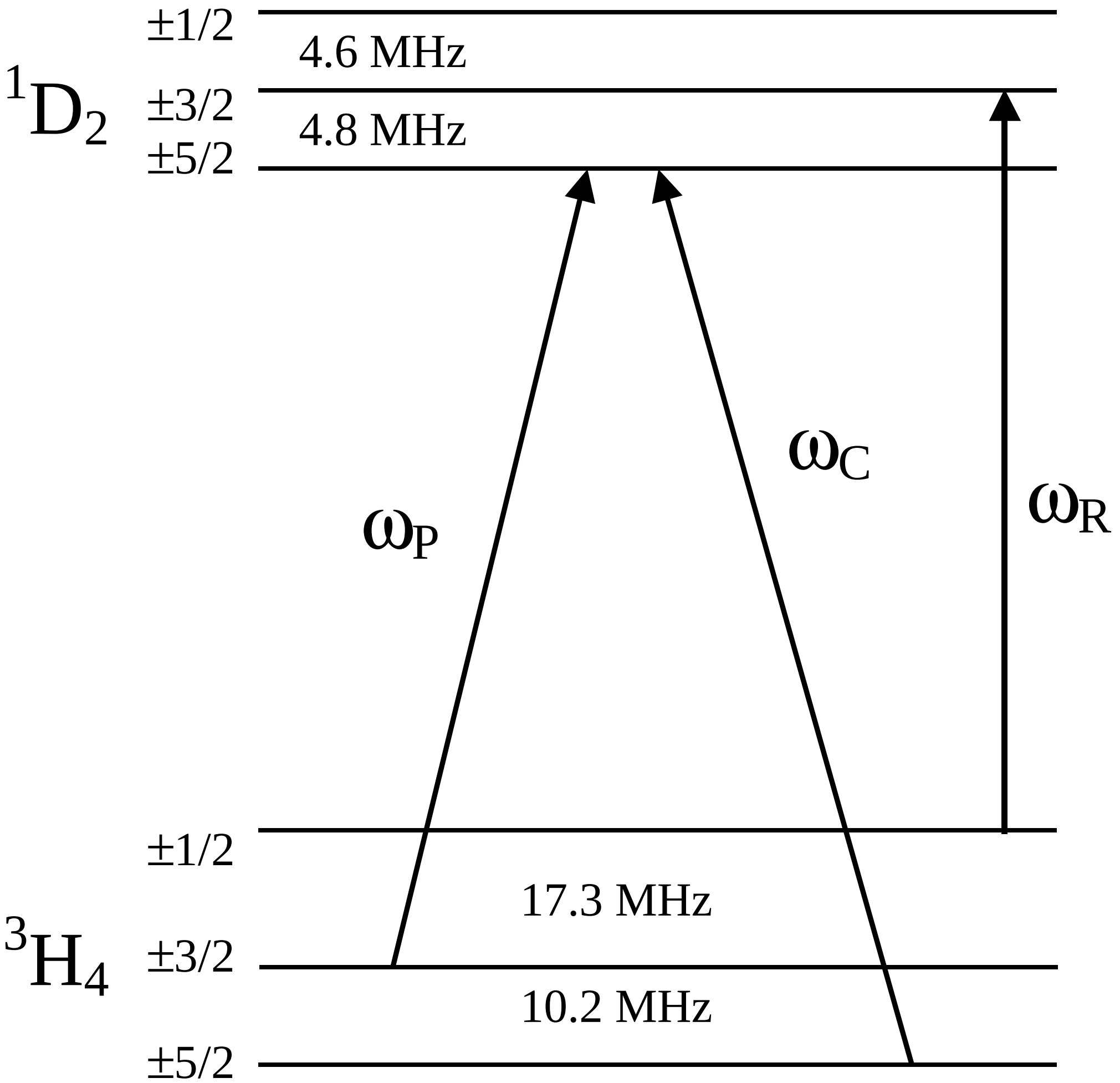

**Figure 1.** Energy level diagram of Pr:YSO. $\omega_C$, $\omega_R$, and $\omega_P$ are frequencies of the coupling, the repump, and the probe fields, respectively.

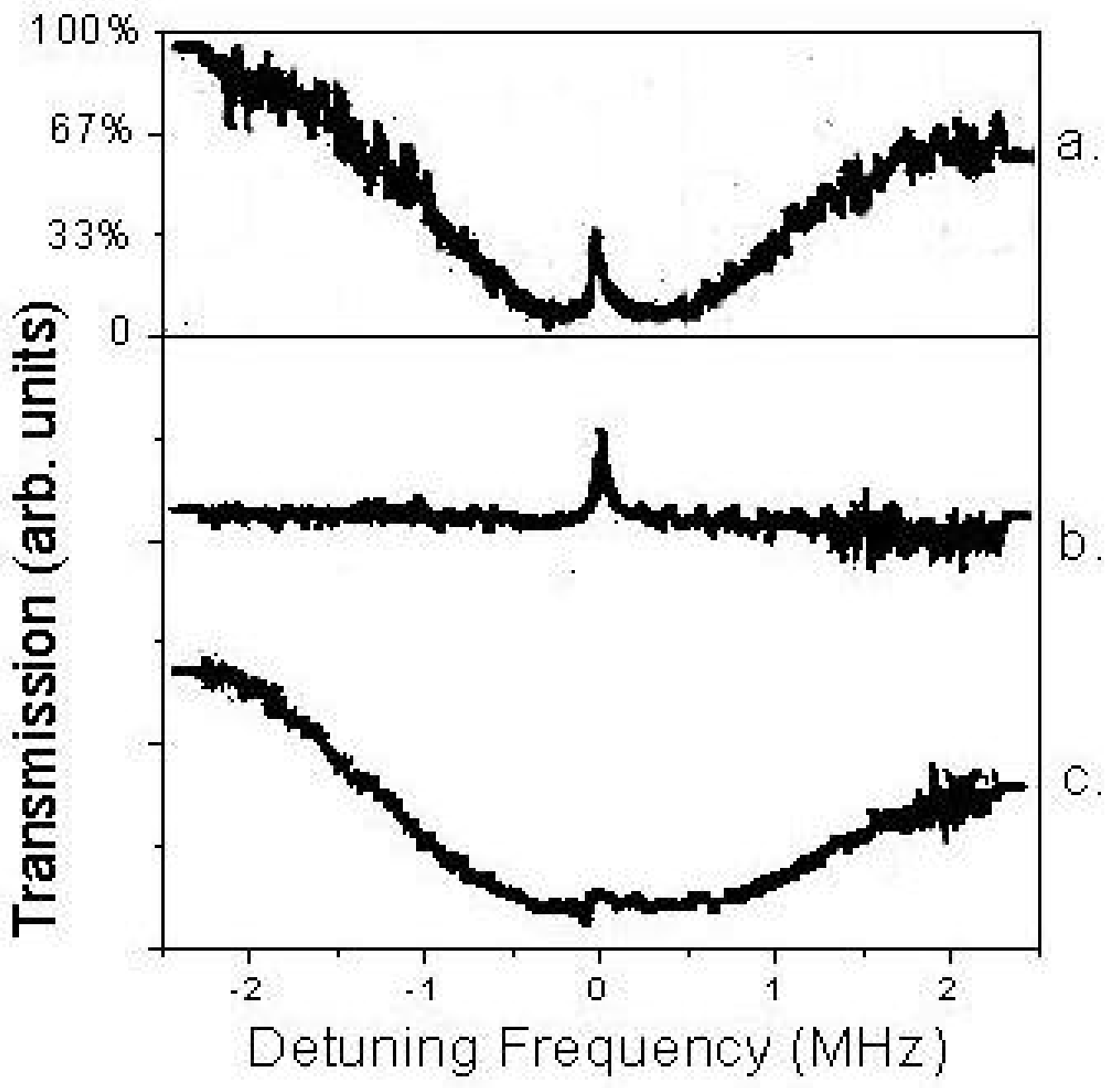

**Figure 2.** The absorption spectra of the probe beam at 5.5 K for a coupling field intensity of 105 W/cm$^2$. (a) spectrum recorded by phase in-sensitive lock-in detection; (b) spectrum recorded by phase sensitive lock-in detection with the phase adjusted to suppress the anti-hole absorption; (c) the spectrum (b) recorded with the phase adjusted to suppress the EIT peak. The modulation frequency of the probe beam was 6 kHz in for all three traces.

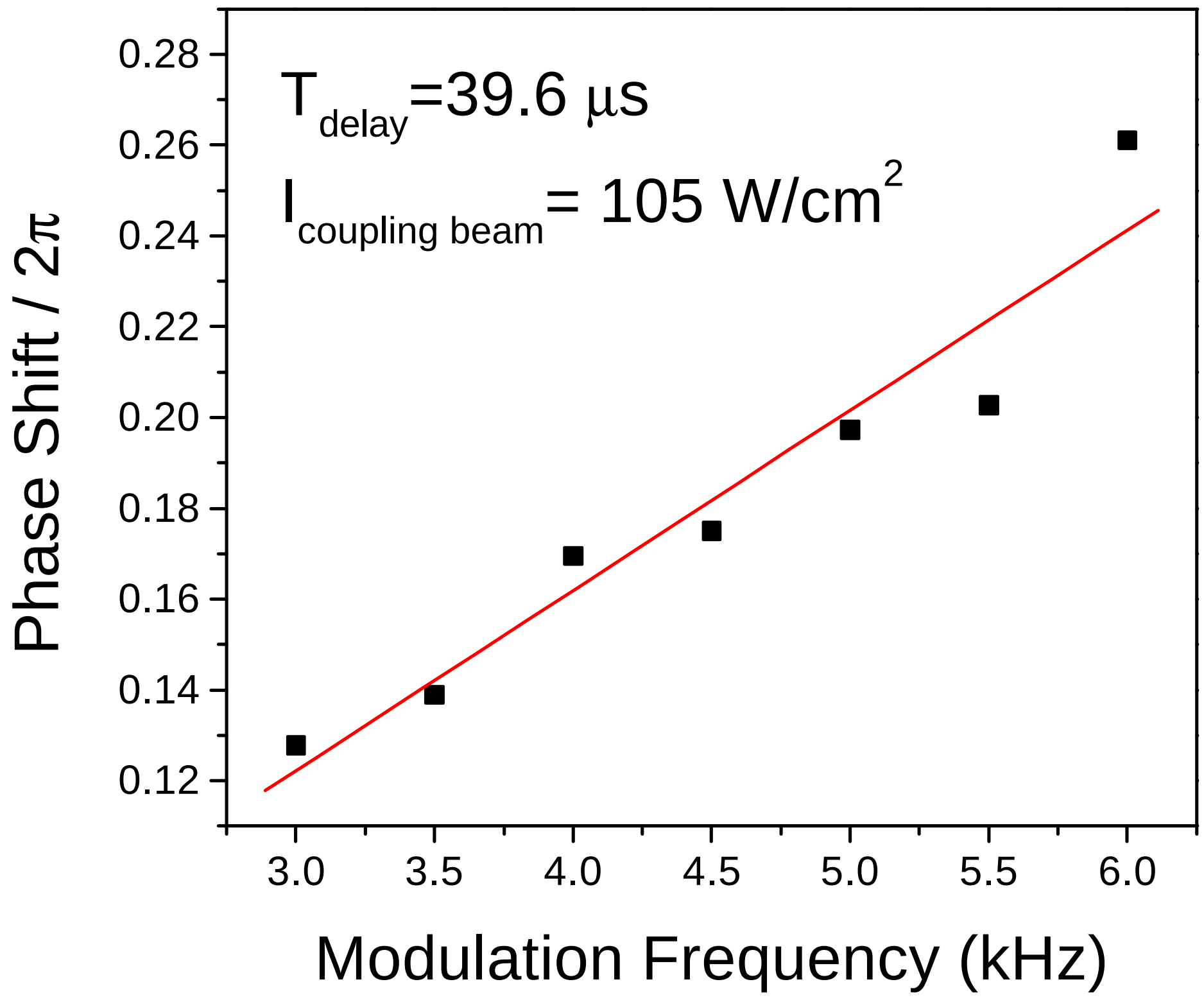

**Figure 3.** Phase shift between the transmitted probe signal with and without the coupling field vs. modulation frequency. The delay time was found to be 39.6 μs from the linear fit of the data.

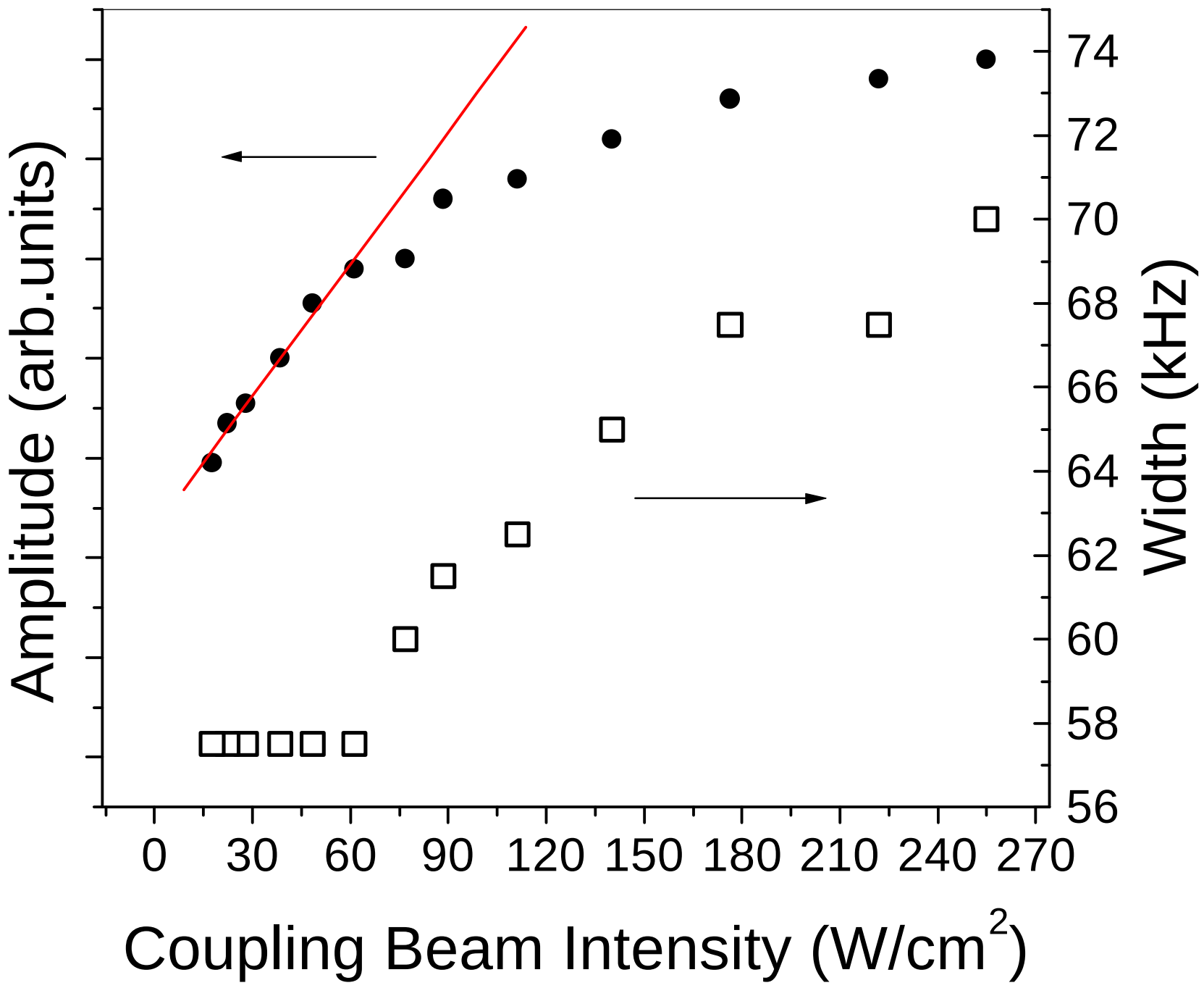

**Figure 4.** The EIT peak amplitude (closed squares) and width (open circles) vs. coupling beam intensity.

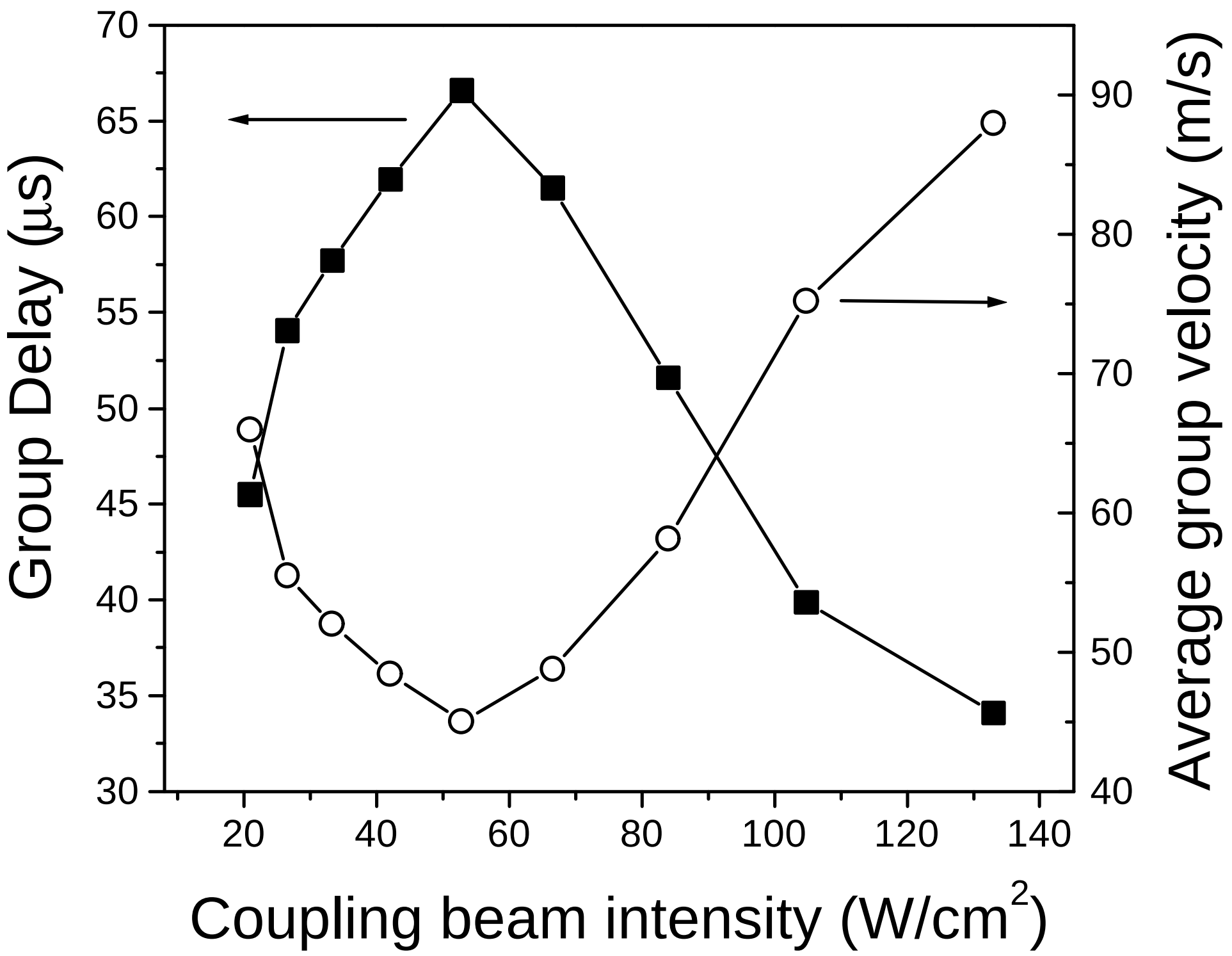

**Figure 5.** Measured group delays (solid squares) and deduced group velocity (open circles) vs. the intensity of the coupling field.